\documentclass[conference]{IEEEtran}
\usepackage{cite}
\ifCLASSINFOpdf
  \usepackage[pdftex]{graphicx}
 \DeclareGraphicsExtensions{.pdf,.jpeg,.png}
\else
 \usepackage[dvips]{graphicx}
 \DeclareGraphicsExtensions{.eps}
\fi
\usepackage{amsmath}
\usepackage{multirow}
\usepackage{array}
\ifCLASSOPTIONcompsoc
 \usepackage[caption=false,font=normalsize,labelfont=sf,textfont=sf]{subfig}
\else
\usepackage[caption=false,font=footnotesize]{subfig}
\fi
\graphicspath{{./figures/}}

\usepackage{url}
\usepackage{array}
\usepackage{algorithm}
\usepackage{algorithmic}
\usepackage{color}
\usepackage{xcolor, soul}
\usepackage{multirow}
\sethlcolor{yellow}
\usepackage{amsmath}
\IEEEoverridecommandlockouts
\begin{document}
\title{Joint Network Coding and Machine Learning for Error-prone Wireless Broadcast$^1$}

\author{\IEEEauthorblockN{Dong Nguyen\IEEEauthorrefmark{3}, 
Canh Nguyen\IEEEauthorrefmark{1},
Thuan Duong-Ba\IEEEauthorrefmark{2},
Hung Nguyen\IEEEauthorrefmark{3},
Anh Nguyen\IEEEauthorrefmark{3}, and
Tuan Tran$^2$\IEEEauthorrefmark{4}
 }
 \IEEEauthorblockA{\IEEEauthorrefmark{1}FPT University, Hanoi, Vietnam\\
Email: canhnht1709@gmail.com}
\IEEEauthorblockA{\IEEEauthorrefmark{2}Hanoi University of Science and Technology,
Hanoi, Vietnam\\
Email: {thuan.duongbahong@hust.edu.vn}}
\IEEEauthorblockA{\IEEEauthorrefmark{3}Saolasoft Inc., CO 880211, USA\\
Email: {\{dnguyen, lnguyen, anguyen\}@saolasoft.com}}
\IEEEauthorblockA{\IEEEauthorrefmark{4}Sullivan University, KY 40205,  USA\\
Email: ttran@sullivan.edu}
\thanks{${}^1$This research will be presented in IEEE-CCWC 2017}
\thanks{${}^2$Corresponding author}
\vspace{-0.4in}
}

\maketitle
\begin{abstract}

Reliable broadcasting data to multiple receivers over lossy wireless channels is challenging due to the heterogeneity of the wireless link conditions. Automatic Repeat-reQuest (ARQ) based retransmission schemes are bandwidth inefficient due to data duplication at receivers. Network coding (NC) has been shown to be a promising technique for improving network bandwidth efficiency by combining multiple lost data packets for retransmission. However, it is challenging to accurately determine which lost packets should be combined together due to disrupted feedback channels. This paper proposes an adaptive data encoding scheme at the transmitter by joining network coding and machine learning (NCML) for retransmission of lost packets.
Our proposed NCML extracts the important features from historical feedback signals received by the transmitter to train a classifier. The constructed classifier is then used to predict states of transmitted data packets at different receivers based on their corrupted feedback signals for effective data mixing. We have conducted extensive simulations to collaborate the efficiency of our proposed approach. The simulation results show that our machine learning algorithm can be trained efficiently and accurately. The simulation results show that on average the proposed NCML can correctly classify 90$\%$ of the states of transmitted data packets at different receivers. It achieves significant bandwidth gain compared with the ARQ and NC based schemes in different transmission terrains, power levels, and the distances between the transmitter and receivers.

\end{abstract}
\IEEEpeerreviewmaketitle
\section{Introduction}
Reliable broadcast is a mechanism for disseminating identical information from one source to many receivers. It is widely used in many applications ranging from satellite communications to wireless mobile ad hoc and sensor networks. For instance, a cluster head may want to reliably broadcast information to hundreds of sensors in its cluster for new software update. In such a scenario, every time the source node sends a data packet to the receivers, a feedback signal is used to inform the source whether it receives the data packet successfully or not via acknowledgement (ACK) and negative-acknowledgement (NAK), respectively. When the communication channels are lossy, ACK or NAK data can be corrupted at the source. As a result, the source node might retransmit redundant data, e.g., packets have been received successfully at the receivers. These unnecessary retransmissions not only reduce transmission bandwidth efficiency but also quickly drain the limited energy of the sensors.



Currently, Forward Error Correction (FEC) in conjunction with Automatic Repeat-reQuest (ARQ) is used to cope with packet errors or losses \cite{Nafaa2000}. Using FEC, the transmitter adds redundant information to transmitted data packets which allows the receivers to detect and correct some of corrupted bits of the received packets. When number of corrupted bits is greater than the correction capability of the FEC codes, transmitter needs to retransmit the corrupted data packet. These approaches work well in small size networks with good channel condition. However, in large networks with the existence of high interference, these approaches are bandwidth inefficient due to redundant retransmission. That will significantly reduce the quality of service of the network.


Recently, network coding \cite{Ahlswede2000} has been shown to be a promising technique for improving the network bandwidth efficiency. Using NC, the transmitter combines the lost data packets across multiple receivers for retransmission\cite{nguyen2007 ,hou2015}. In this way, with a single retransmission of network coded packet, multiple receivers can recover their lost data simultaneously, resulting in significant bandwidth efficiency improvement. However, the existing NC-based approaches are usually assumed that the transmitter instantly knows states of transmitted data packets at the receivers \cite{nguyen2009, tran2016} to determine which packets should be combined for retransmissions. However, such an assumption does not hold in practical transmissions where both transmission and feedback links are subject to errors. Some approaches, e.g., \cite{nguyen2007m,nguyen2011}, applied the framework of Markov Decision Process at the transmitter to determine optimal packet combination via statistics of link conditions. These approaches, however, suffer the curse of dimensionality when number of receivers and transmitted data packets increases.



In this paper, we propose a new approach by combining machine learning with network coding for efficient bandwidth usage. In our approach, the transmitter exploits the historical feedback signals from receivers to train a classifier which is then used to predict the states of new transmitted data packets. The transmitter then uses the predicted outcomes of the classifier to determine its optimal data mixing for retransmission. The proposed machine learning based network coding scheme (NC-ML) is efficient and data-driven without assumption on the knowledge of the transmission channel statistics. Our simulation results show that NC-ML can achieve very high classification accuracy of the states of transmitted data packets and its accuracy improves over time when more data is collected for training our classifier.


The rest of the paper is organized as follows. In Section \ref{sec:related}, we discuss some recently published network coding and machine learning papers in wireless networks. In Section \ref{sec:model}, we describe our system model and background on network coding and supervised machine learning. We then describe in detail our proposed NC-ML approach in \ref{sec:ncml}. In Section \ref{sect:simulation}, we present our extensive simulation results and discussion. Finally, we conclude our paper in Section \ref{sec:con}.
\begin{figure}[!t]
\centering
\includegraphics[width=2.0in]{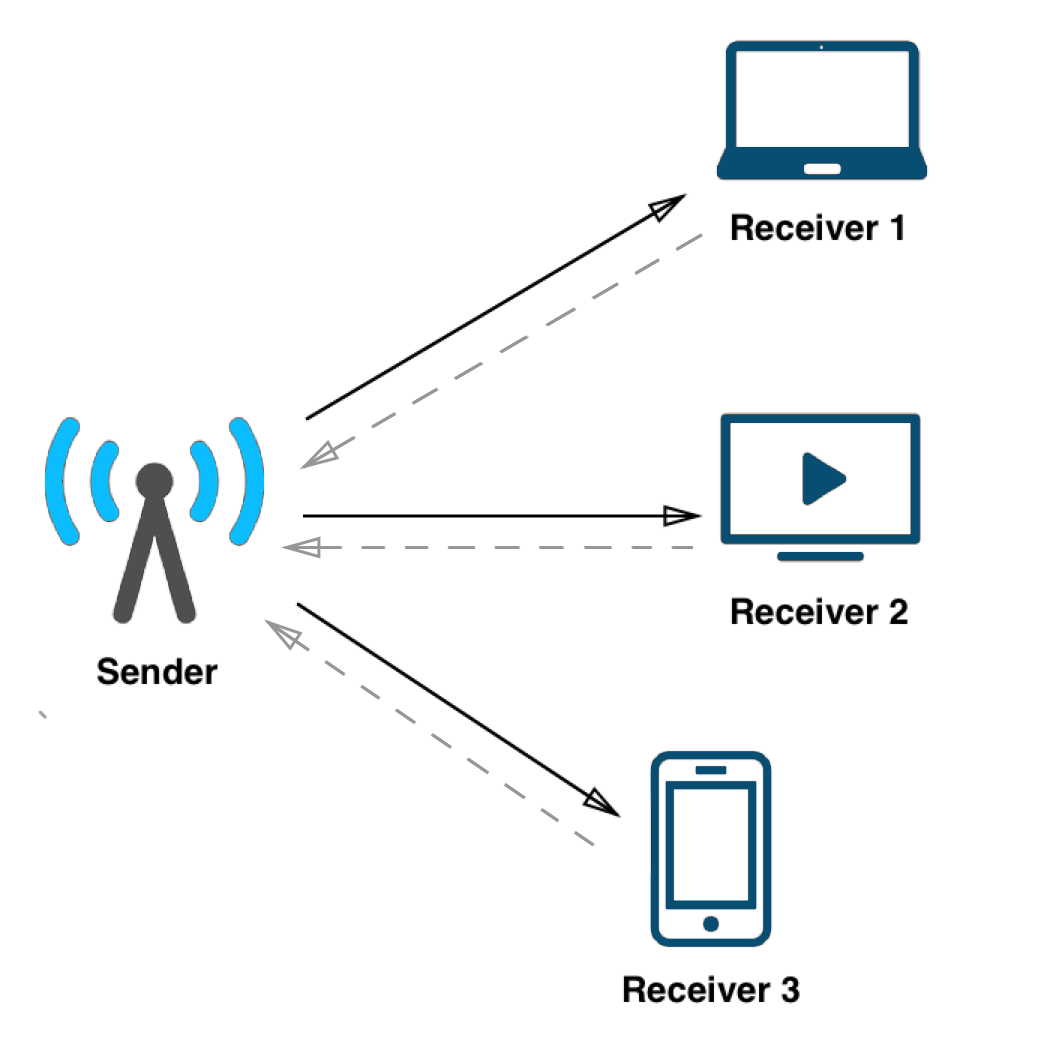}
\caption{A typical setting for reliable wireless broadcast in which a sender sends a stream of packets to multiple receivers and the receivers send feedback to the sender to notify the status of received packets.}
\label{transmission_with_feedback}
\end{figure}


\section{ Related Work }
\label{sec:related}
Network coding has been shown to be an effective technique for sending data both for multicast and broadcast scenario, particularly in wireless broadcast networks \cite{widmer2005, tian2004, katti2006, nguyen2007, nguyen2007m}. In \cite{widmer2005}, the authors consider the problem of wireless broadcasting in ad-hoc network with the use of network coding that helps to reduce the energy consumption significantly. Katti et al. \cite{katti2006} and Nguyen et al. \cite{nguyen2007} show that network coding can improve the channel efficiency both in theory and practical networks. In addition, the wireless network coding schemes are applied in scheduling for video and multimedia streaming over the erroneous network to increase the multimedia transmission quality\cite{nguyen2007m, tran2016}.

Recently, many studies have been conducted on machine learning used in the dynamic wireless networking environment to improve the transmission throughput. \cite{Alsheikh} lists comprehensive review on how machine learning techniques can be used to provide the solution for maximizing the resource utilization and prolonging the network lifespan of wireless sensor networks. In another study \cite{rico2014learning}, the authors illustrate a data-driven approach to link adaptation, where a machine learning classifier is employed to determine the optimal modulation and coding scheme for multiple input-multiple output (MIMO) systems.  Kotobi et al. \cite{kotobi2015data} propose the data-mining cognitive radio using learning techniques for decision making and improving the performance of a wireless network. By mining the dataset, the algorithms create more efficient cognitive radio networks. In addition, \cite{huang2014wireless} introduces frequency pattern mining to efficiently allocate wireless spectrum between licensed and unlicensed users. In \cite{alsheikh2014machine}, the authors present a review of using machine learning algorithms for wireless sensor networks to adapt the environment changes over time, from that, to increase the network efficiency and save the limited resources of the networks.

Differently, in this paper, we propose a new method by utilizing machine learning algorithms to predict states of transmitted data packets and adaptively use network coding for retransmission. Our proposed approach is efficient and accurate without any assumption on the knowledge of transmission channel statistics. We will describe in detail our system model and the proposed approach in the next sections.



\section {System Model}
\label{sec:model}

We consider reliable wireless broadcasting for multiple receivers in lossy networks. In this setting, the transmitter wishes to reliably broadcast data to several receivers with different channel conditions. Our system model and related background are described in the next subsections.

\begin{figure}[!t]
\centering
\includegraphics[width=3.5in]{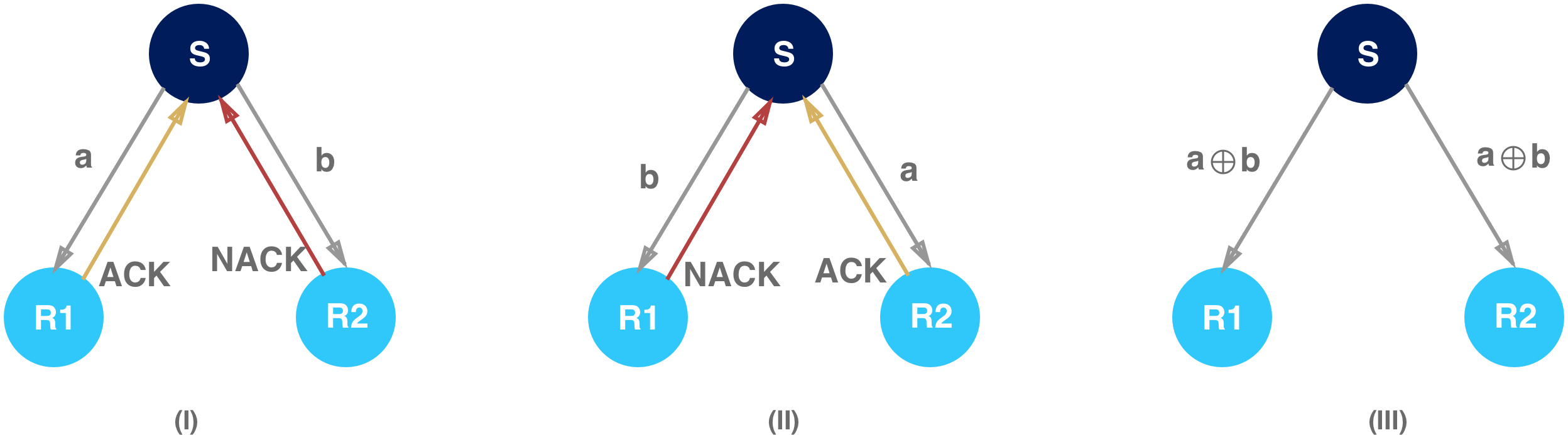}
\caption{A network coding model for wireless broadcast with feedback: (i) S broadcasts packet a to R1 and R2, R2 does not receive it correctly, (ii) S broadcasts packet b to R1 and R2, R1 does not receive it correctly, (iii) S broadcasts the combined packet $a \oplus b$ so that R1, R2 can decode lost packets.}
\label{basic_netcode}
\end{figure}

\subsection{Reliable Wireless Broadcasting over Erroneous Links}

We consider data broadcasting in lossy wireless networks where the transmitter wants to reliably broadcast data to several receivers as illustrated in Figure \ref{transmission_with_feedback}. Every transmitted packet from the transmitter to receiver $R_i$ is subject to an error probability $p_i$. To ensure reliable data transmission, feedback signals, i.e., ACK/NAK, are sent from receivers to inform the transmitter the states of the transmitted packet. In our paper, we assume that the feedback links are also subject to errors. Whenever the transmitter receives an NAK signal, it will retransmit the lost data packet to the signaling receiver until it receives the packet successfully. When all receivers obtain the transmitted packet successfully, the transmitter will proceed to the next data packet. The process is repeated until all data packets are successfully delivered.

\subsection{Wireless Network Coding}

Network coding has been used to improve bandwidth efficiency of lossy wireless networks \cite{nguyen2009}. To illustrate how NC is used to improve the network bandwidth, we provide a simple example in Figure \ref{basic_netcode}. In this example, a transmitter, e.g., an access point or a base station, wishes to reliably broadcast a stream of data packets to two receiver R1 and R2. For the sake of simplicity, we assume that the transmitter has two packets a and b for transmission. Due to erroneous transmission links, we assume that packet $a$ is received correctly at R1 while lost or received incorrectly at R2. In the second transmission, packet $b$ is lost at R1 while successful at R2.

In the NC-based retransmission, the transmitter won't retransmit lost data immediately. Instead, the transmitter will mark lost packets at different receivers and consider data combination for retransmission. Particularly, the transmitter broads a combined packet $c = a \oplus b$ to all receivers. Upon receiving the coded data packet successfully, both R1 and R2 can recover their wanted data by using $a \oplus (a \oplus b)$ and $b \oplus (a \oplus b)$, respectively. In this example, by using NC, the number of transmissions is reduced from 4 to 3, thus, resulting in about $33\%$ bandwidth gain.

\begin{figure}
\centering
\includegraphics[width=3.2in]{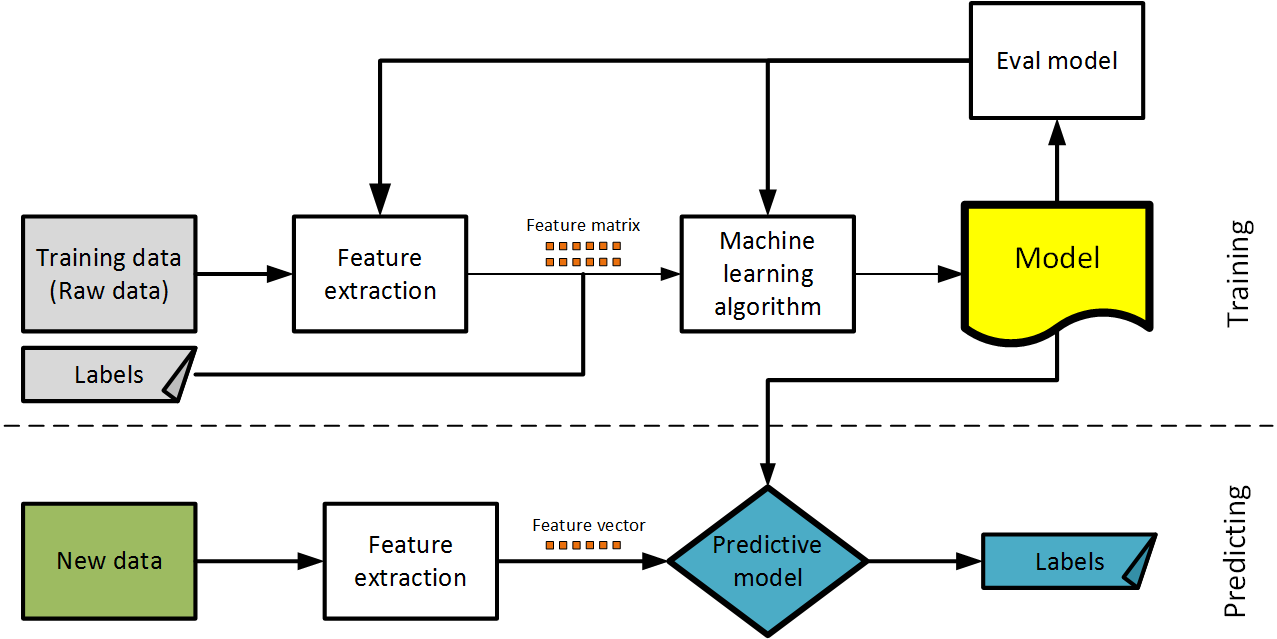}
\caption{A flowchart of a supervised machine learning model}
\label{supervised_ml_model}
\vspace{-0.2in}
\end{figure}

\subsection{Machine Learning for Classification}
\label{ml_classification}

Machine Learning has been widely used to solve several challenges in many fields, including wireless sensor networks. In this subsection, we present some of the classification techniques which are used to identify the transmitted feedback from the receiver. For the sake of completeness, we provide a brief introduction of supervised machine learning for classification here while its detail can be found in \cite{Alpaydin2014} .

Figure \ref{supervised_ml_model} illustrates a typical workflow of a supervised learning algorithm for classification. In the first step, raw training data is cleaned and fed into feature extraction to select only useful features from all available ones. Next, labels or known class and extracted features are then passed to the training phase where machine learning algorithms are used to identify a good model which map inputs to desired outputs. The evaluation phase provides feedback to the feature extraction and learning phases for adjustment to improve model accuracy.The training process is repeated until a desired accuracy level achieved. Once a model is constructed, it is then used to predict label of new data. We will utilize the framework of supervised learning to train our classifier to predict the state of feedback signals in the next section.

\vspace{-0.05in}
\section{The Proposed Machine Learning Based Network Coding}
\label{sec:ncml}

\begin{figure}
\centering
\includegraphics[width=2.8in]{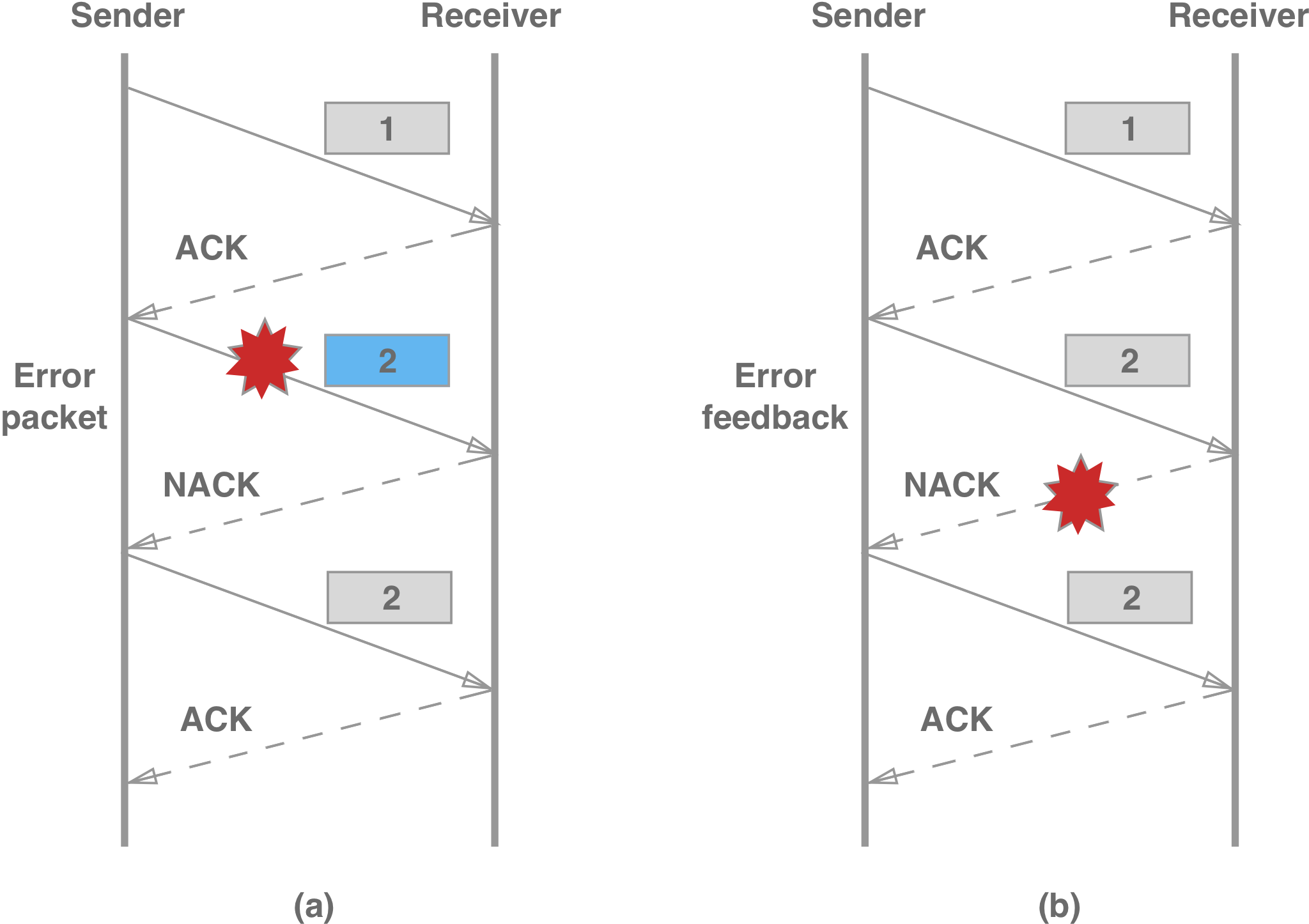}
\caption{Inefficiency of error-prone feedback: (a) the sender correctly retransmit the error packet; (b) the sender sends duplicate packet 2 due to the error feedback.}
\label{error_feedback_example}
\vspace{-0.2in}
\end{figure}

\subsection{The Impact of Error-Prone Feedbacks}
To provide reliable wireless broadcast in lossy networks, feedback signals are used to request retransmission of a lost data packet. Despite of using higher redundancy, the feedback signals are also subject to error. Network bandwidth efficiency will decrease significantly when feedback signals are misclassified. Figure \ref{error_feedback_example} illustrates an example of inefficient bandwidth usage when feedback signal is incorrectly estimated.

In Figure \ref{error_feedback_example}(a), packet 2 is corrupted during transmission, thus, an NAK is sent by the receiver over the feedback channel to inform the transmitter for retransmission. Upon receiving of the feedback signal, the transmitter retransmits packet 2. However, due to the error in the feedback channels, feedback signals are subject to error as well. Without any mechanism to detect the states of a transmitted data packet at the receivers, the transmitter might retransmit redundant data packet, resulting in inefficient bandwidth usage. Figure  \ref{error_feedback_example}(b) shows an example of impact of error feedback on transmission bandwidth. In this case, packet 2 is actually received successfully at the receiver; however, due to feedback error, the transmitter still retransmits packet 2. This inefficiency of retransmission can render more overhead in a broadcast channel in which one error on particular feedback can initiate a retransmission to all other receivers in the networks. The erroneous feedback not only decreases bandwidth efficiency but also quickly drains out limited power of wireless receivers.

\subsection{The Proposed Machine Learning Based Network Coding (NCML)}

In this subsection, we will describe our proposed machine learning based network coding (NCML) approach for reliable data broadcast in lossy wireless networks. Our NCML illustrated in Figure \ref{feature_data_ml} consists of three main steps:

\begin{itemize}
\item {\bf Collecting Feedback Data}: In the first step, historical feedback data is collected based on realistic simulation of data broadcast networks. In particular, we simulate data transmission in different settings including transmission range, transmit power, proximity environment, modulation, etc. to gather our feedback signal data.
\begin{itemize}
\item {\it Transmission Environments}: To simulate transmission different environments, we adopt the link models proposed in \cite{erceg1999}
\begin{eqnarray}
\label{eqn:pl}
PL&= [A + 10(a - bh_b + c/h_b)log_{10}(d/d_0)] + \nonumber \\
&[10x\sigma_\gamma log_{10}(d/d_0) + y\mu_\sigma + yz\sigma_\sigma],
\end{eqnarray}
where $A = 20log_{10}(4\pi d_0/\lambda)$ denotes the fixed free-space path loss at reference distance $d_0$ and wave length $\lambda$; $a, b,$ $c$ and $\sigma_{\gamma}$ are data-derived constants for each terrain category; $h_b$ denotes transmitter antenna height in meters; $d$ is distance between transmitter and receiver; $x$, $y$, and $z$ denote a zero-mean Gaussian variable of unit standard deviation $N[0,1]$; $\mu_{\sigma}$ and $\sigma_{\sigma}$ are are both data-derived constants for each terrain. Typically, the first part of Equation (\ref{eqn:pl}) denotes median the path loss at distance $d$ depending on a specific terrain, base antenna height and transmission distance while the second part represents a zero-mean random variation about that median.
\item {\it Data Generation}: Simulation program is run several times to collect data of different transmission configurations. The states of feedback signals are labeled based on their decoded data (i.e., ACK or NAK). At the training step, we discard all corrupted feedback signals which cannot be decoded.
\end{itemize}
\item {\bf Supervised Learning Classifier for Erroneous Feedbacks}: In the first step, the transmitter constructs a classifier for predicting the states of data packet received at different receivers based on the historical feedback data. It includes the following components:
\begin{itemize}
\item {\it Historical Feedback Data}: This component collects and stores the past data of feedback signals from different receivers. The states of feedback signals are labeled based on only the ACK/NAK signals which don't have any errors. In other words, only successfully received ACKs/NAKs at the transmitter are used for labeling the states of the transmitted data packets.
\item {\it Feature Extraction and Selection}: Obviously, there are many features of the communication links between the transmitter and receivers can be utilized for constructing our classifier. However, some of these features are correlated. This step is to extract and select features of feedback signals for training classifier in the next step. Table I shows list of our extracted features and their units used in our training phase.
\item {\it Training Classifier}: Our classifier is constructed by using features extracted from the Feature Extraction and Selection step. The collected dataset is first partitioned into training and validation datasets. The training dataset is used to construct our classifier and the validation dataset is used to evaluate the constructed classifier. This training step is carried out iteratively on different combination of the features. The classifier with minimum classification error on the validation dataset is selected as our final classifier. Algorithm \ref{alg:training_pseudo} describes pseudocode of the training process.

\begin{algorithm}[H]
%
%
%
%
%
\begin{algorithmic}[1]
 \renewcommand{\algorithmicrequire}{\textbf{Input:}}
 \renewcommand{\algorithmicensure}{\textbf{Output:}}
 \REQUIRE Historical feedback dataset with features and labels
 \ENSURE  The best classifier (i.e., minimum classification error)
 \\ \textit{Initialization} : $E \leftarrow$ collected feedback dataset; $\Sigma \leftarrow \{\text{set of machine learning techniques}\} $; $\Delta \leftarrow\{\text{set of all feedback signal features}\}$; $n \leftarrow |\Delta|$; $P_0 \leftarrow \emptyset$
 \FOR {$T \in \Sigma$}
  	\STATE {\textit{best} $\leftarrow$ CHOOSE$\_$ATTRIBUTE\{$\Delta, E$, method = $T$\}}
	\STATE{$P \leftarrow$ TRAIN$\_$CLASSIFIER(\textit{best}, data = $E_{training}$, method = $T$)}
	\IF{(EVALUATE($P$, data = $E_{Validation}$) $>$ EVALUATE($P_0$, data = $E_{Validation}$))}
		\STATE {$P_0 \leftarrow P$}
	\ENDIF
 \ENDFOR
 \RETURN $P_0$
 \end{algorithmic}
 \caption{Pseudocode of NCML Classifier}
 \label{alg:training_pseudo}
\end{algorithm}
\end{itemize}

\begin{table}
\center
\caption{Feedback Signal Parameters}
\begin{tabular}{ |c|l| p {5cm} |}
\hline
\bfseries No. & \bfseries Features & \bfseries Description \\
\hline
1& distance (m) & Estimated distance between the transmitter and receivers\\
\hline
2 & noise (dBm) & Noise level\\
\hline
3 & terrains & Types of transmission environments (terrains)\\
\hline
4 & snr (dBm) & Signal to noise ratio of feedback signal\\
\hline
5 & rx (dBm) & Received power of feedback signal at the transmitter\\
\hline
6 & mod & Types of modulation \\
\hline
\end{tabular}
\end{table}
\vspace{0.05cm}




\item {\bf Machine Learning Based Network Coding Retransmission}: In the third step, NCML performs network coding based data mixing for retransmissions. Our transmission includes two phases: transmission and retransmission phases. In the transmission phase, the transmitter keeps sending data packets to the receivers and collects feedback signals. These feedback signals are fed to our constructed classifier to determine the states of the transmitted data packets at different receivers. The output of the classifier is used to create a packet state map of the transmitted data packets at different receivers for all data packets transmitted in transmission phase. After the transmission phase, the transmitter will consider lost packets for retransmission. Based on the constructed packet state map, the transmitter will determine which lost packets are combined together for retransmission.


\begin{figure}
\centering
\includegraphics[width=3in]{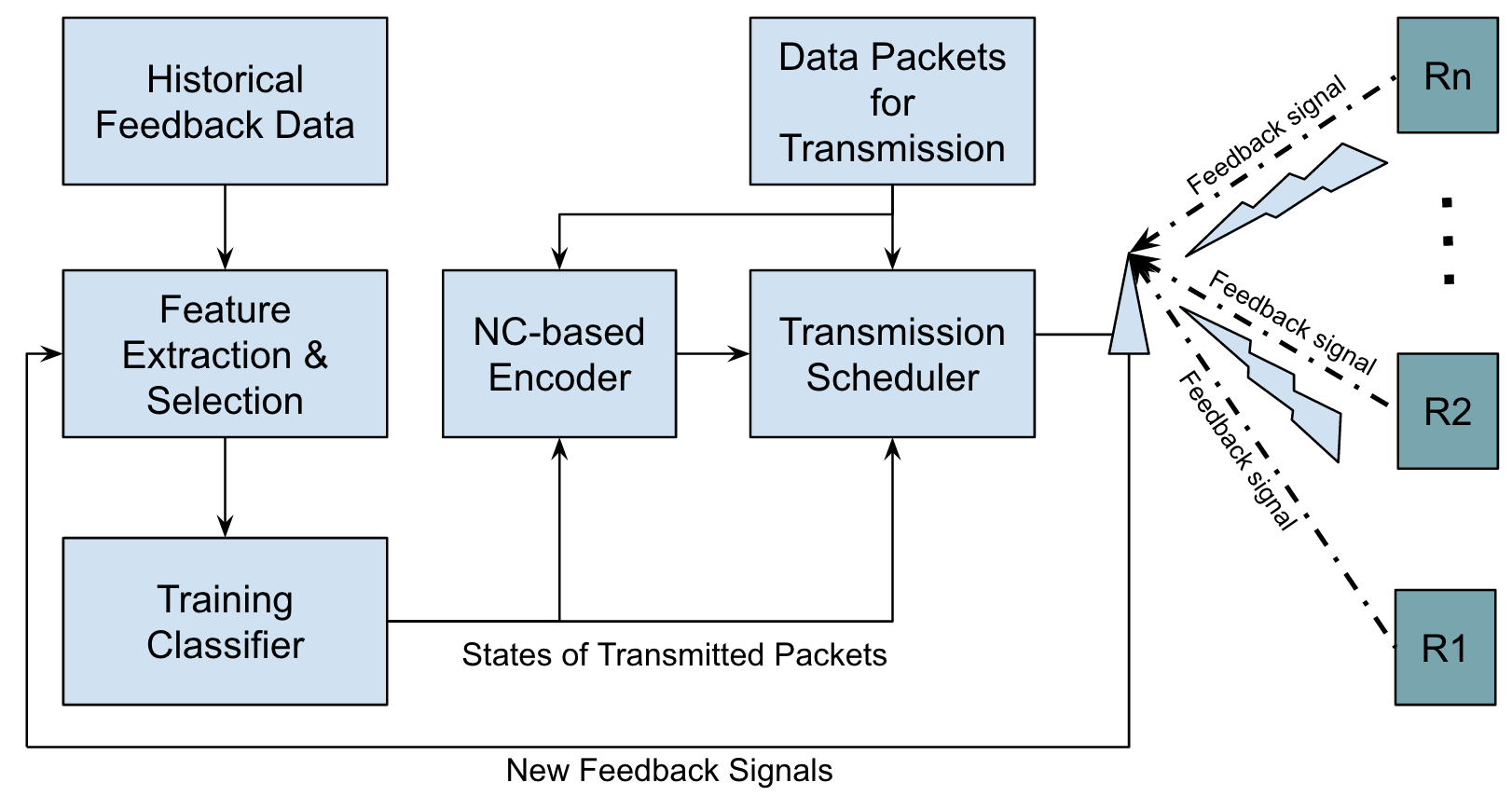}\\
\caption{Block diagram of NCML}
\label{feature_data_ml}
\end{figure}

\end{itemize}

\section{Simulation Results and Discussion}
\label{sect:simulation}

\subsection{Simulation Setting and Transmission Schemes}
We set up a basic simulation setting of a WiMAX based wireless broadcast scenario in which there are one transmitter and multiple receivers. We assume that data is always available for transmission at the transmitter. Transmission is divided into transmission and retransmission phases. In the transmission phase, data packets are transmitted to all receivers and feedback signals are collected. In our simulation, both transmission and feedback signals are subject to errors due to interference and fading in the wireless networks. When a data packet is lost at a receiver, the transmitter will retransmit it until it is received successfully. In our simulation, we consider the following types of terrains:

\begin{table}[]
\centering
\caption{Parameters of Simulation Environments}
\tiny
\label{my-label}
\begin{tabular}{|c|c|c|c|}
\hline
\multirow{2}{*}{MODEL PARAMETER} & \multicolumn{3}{c|}{TERRAIN CATEGORY} \\ \cline{2-4}
 & \begin{tabular}[c]{@{}c@{}}Terrain 1\\ (Hilly/Moderate-to-\\ Heavy Tree Density)\end{tabular} & \begin{tabular}[c]{@{}c@{}}Terrain 2\\ (Hilly/Light/Tree Density \\ or Flat/Moderate-to-\\ Heavy Tree Density)\end{tabular} & \begin{tabular}[c]{@{}c@{}}Terrain 3\\ (Flat/Light Tree \\ Density)\end{tabular} \\ \hline
a & 4.6 & 4 & 3.6 \\ \hline

\begin{tabular}[c]{@{}c@{}}b (in m\textsuperscript{-1})\end{tabular} & 0.0075 & 0.0065 & 0.005 \\ \hline
\begin{tabular}[c]{@{}c@{}}c (in m)\end{tabular} & 12.6 & 17.1 & 20 \\ \hline
$\sigma_\gamma$ & 0.57 & 0.75 & 0.59 \\ \hline
$\mu_\sigma$ & 10.6 & 9.6 & 8.2 \\ \hline
$\sigma_\sigma$ & 2.3 & 3 & 1.6 \\ \hline
\end{tabular}
\end{table}
\begin{itemize}
\item {\bf Path Loss Model}: We simulate suburban areas with different terrain categories \cite{erceg1999} in Table \ref{my-label}.
In our simulation, we also vary the distance between the transmitter and receivers as well. This feature will be exploited in training our model.

%
%

\item {\bf Transmission Schemes}: In our simulation, we consider different transmission schemes for reliable data delivery.

\begin{itemize}
  \item \textbf{Naive re-transmission ARQ (ARQ) Scheme}: In this scheme, if an NAK is received from a receiver, the transmitter retransmits the lost packet until the receiver receives it successfully. We emphasize that in this scheme, the state of the transmitted data packet is purely based on the recovery of feedback signal received at the transmitter. The process is repeated to all receivers until all data packets are delivered successfully to all receivers. This simulates a typical existing reliable data transmission protocol.
  \item \textbf{ARQ with Machine learning (ARQ-ML) Scheme}: In ARQ-ML, the states of a transmitted data packet at different receivers are determined by the constructed classifier. The feedback signals from the receivers are fed through our constructed classifier and its output will be used to determine if the data packet is retransmitted to a receiver or not. The process is repeated until all data packets are delivered successfully to all receivers.
  \item \textbf{Network Coding (NC) Scheme}: In this scheme, the transmitter implements NC for retransmission. Basically, the transmission is divided into two phases. In the first phase, the transmitter sends out all data packets to the receivers and collect feedback signals from the receivers. In the second phase, the transmitter will consider lost data packets for retransmission. The combination is performed adaptively based on the states of retransmission at different receivers. In this scheme, the states of transmitted data packets are based on the recovery of feedback signals from the receivers.
   \item \textbf{Machine Learning Based Network Coding (NC-ML) Scheme}: Similar to NC schemes, the NC-ML scheme also divides the data transmission into transmission and retransmission phases where the transmitter sends out data packets in the first phase and collect states of transmitted data packets at different receivers. The second phase is for retransmission of lost data packets. The difference between NC-ML compared to the NC is that in NC-ML, the states of data packets are predicted by using our constructed classifier based on the feedback signals from the transmitter.

\end{itemize}
\item {\bf Performance Metric}: To compare different transmission schemes, we use the network effective throughput defined as the average number of transmission per data packet per receiver. Mathematically, we define our effective network throughput by
\begin{equation}
\eta = \lim_{N \rightarrow \infty} \frac{\sum_{i=1}^Nn_i}{NMK},
\end{equation}
where $N$ denotes number of simulation trials, $M$ is number data packets for transmission, $K$ and $n_i$ are number of receivers and number of transmission of trial $i$, respectively. The transmission scheme with smallest $\eta$ is the best approach.

\end{itemize}


\begin{figure}[!t]
\centering
\includegraphics[width=3in]{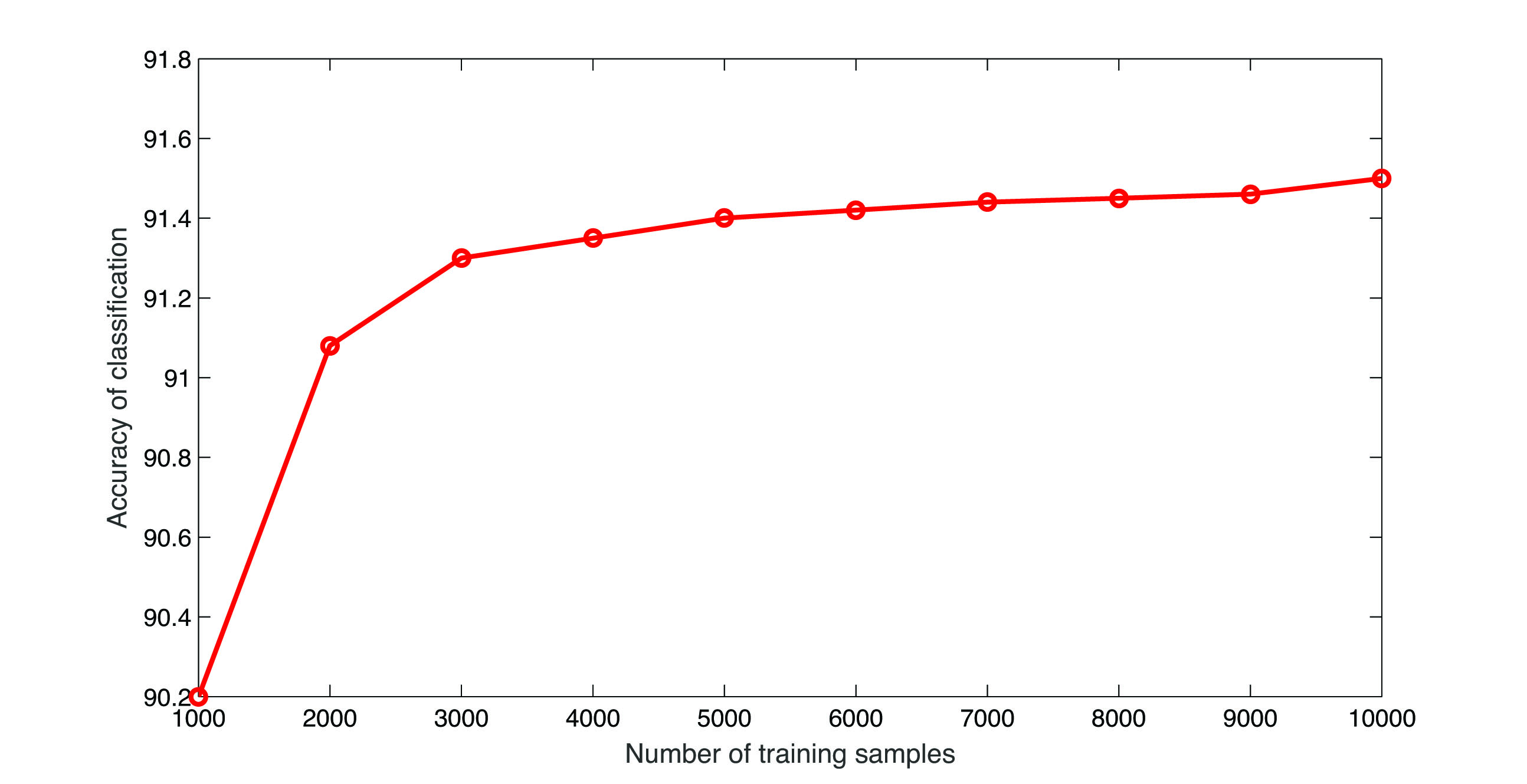}
\caption{Accuracy of machine learning algorithm versus the training size.}
\label{accuracy_vs_datasize}
\end{figure}

\subsection{Results and Discussion}


We first evaluate the performance of our machine learning classifier with different sizes of training dataset in Figure \ref{accuracy_vs_datasize}. As expected, the accuracy of our classifier increases proportionally to the size of the training dataset. This is because when number of training examples increases, the training dataset retains almost all important information of the population which then translates into the classifier, resulting in higher accuracy.

\begin{figure}[!t]
\centering
\includegraphics[width=2.5in]{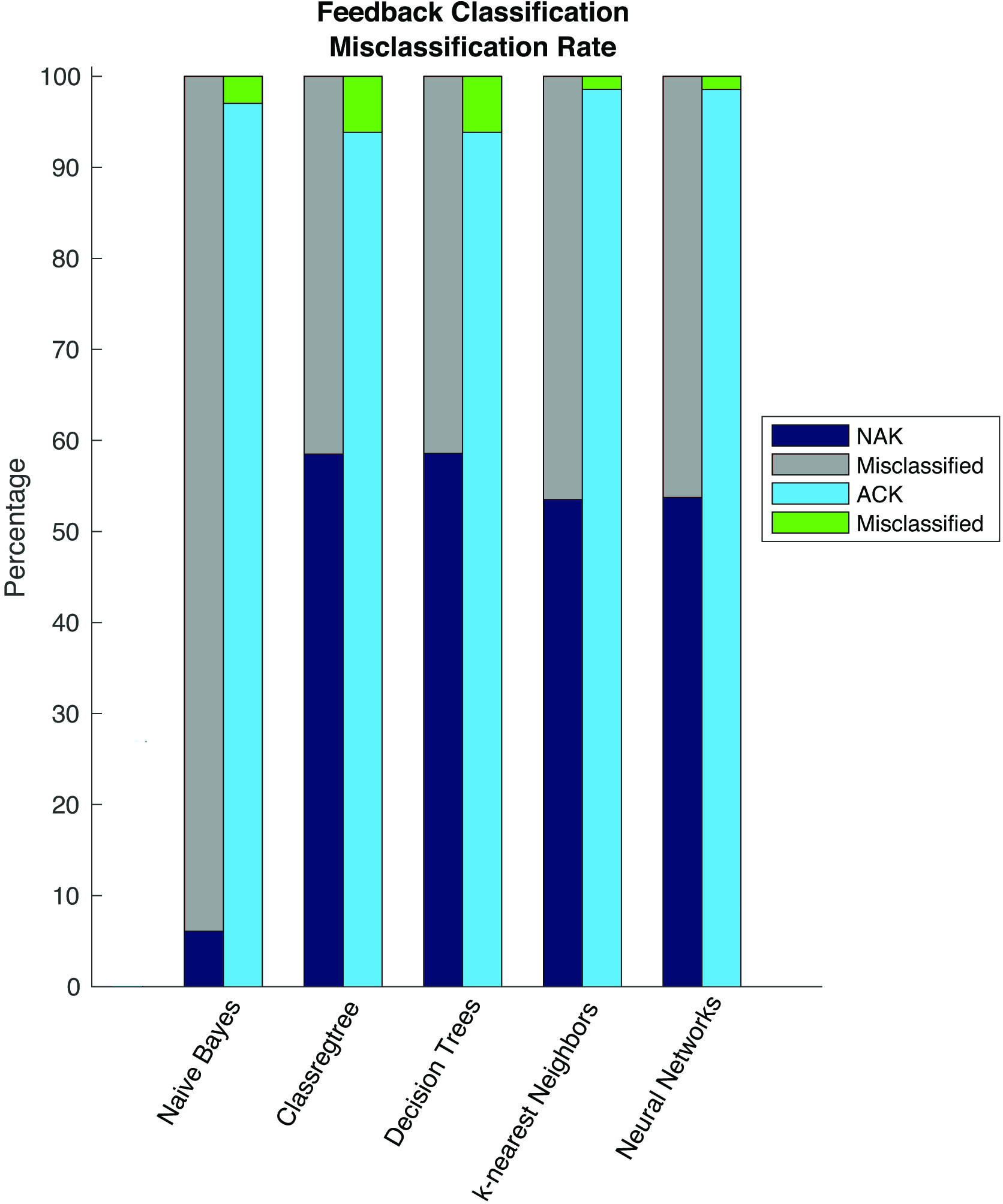}
\caption{Comparison of different machine learning models}
\label{ml_compare}
\vspace{-0.15in}
\end{figure}

Next, we compare the classification accuracy of different machine learning algorithms given the same size of training dataset. As shown in Figure \ref{ml_compare}, the Neural Networks obtained the best performance with an average classification accuracy of $90\%$ while Naive Bayes has the worst performance. This can be explained by the flexibility thanks to multiple layers of the neural networks which can be trained a classifier to fit to non-linear data. However, the high accuracy of the Neural Networks also comes with the longer training time.


\begin{figure}[!t]
\centering
\includegraphics[width=2.7in]{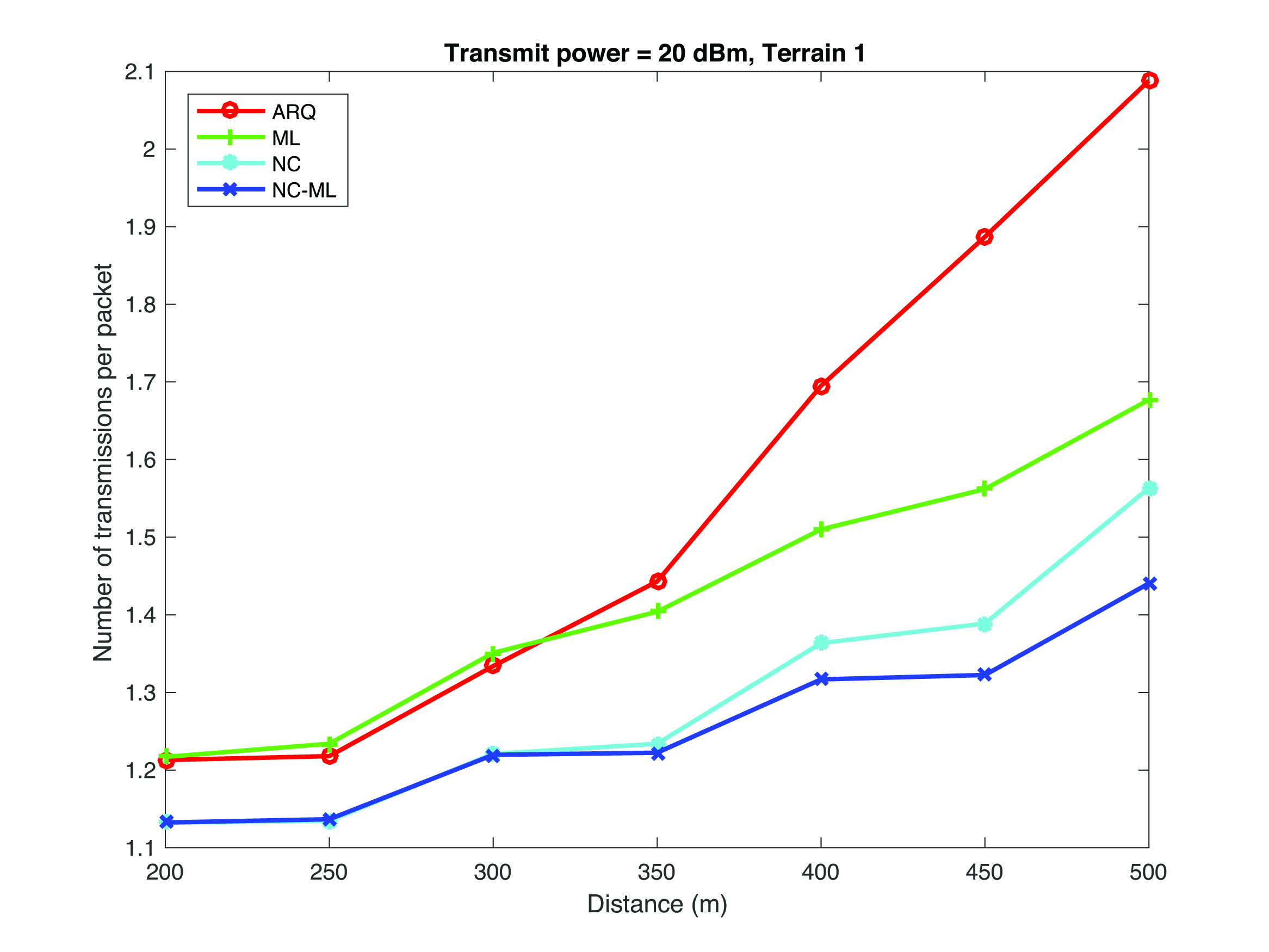}
\caption{Bandwidth efficiency vs. transmission distance}
\label{xdistance_y_transmition_power_20}
\end{figure}

In addition, we evaluate the network bandwidth efficiency of different schemes in Figure \ref{xdistance_y_transmition_power_20}. Particularly, we compare the average number of transmissions per packet of different schemes versus transmission distances. In this simulation, we consider two receivers and both of them have the same distances to the transmitter. Those distances vary from 200 meters to 500 meters while the transmit power is unchanged at 20 (dBm). As expected, the number of transmissions per packet increases with the distance. This is because the received powers at the receivers decrease when distance increases. As a result, more number of packets will be lost during transmission requiring more number of retransmissions. We also observe that ARQ and NC-ML achieve the worst and best performance, respectively. We further observe that in the regime of short distance, i.e., from 200 to 300 meters, the machine learning based schemes, i.e., ML and NC-ML, obtain similar performance compared with their counterparts, i.e., ARQ and NC, respectively. This is because in the range of short distance, most of feedback signals are not corrupted. Therefore, there is not much gain by using machine learning in predicting the states of data packets in the regime. However, the performance gain increases when transmission range is greater than 300 meters. This is because more feedback signals got corrupted and machine learning based schemes bring benefit by predicting correctly the actual states of transmitted data packets at the receivers. As a result, they reduce the number of retransmissions.


\begin{figure}[!t]
\centering
\includegraphics[width=2.8in]{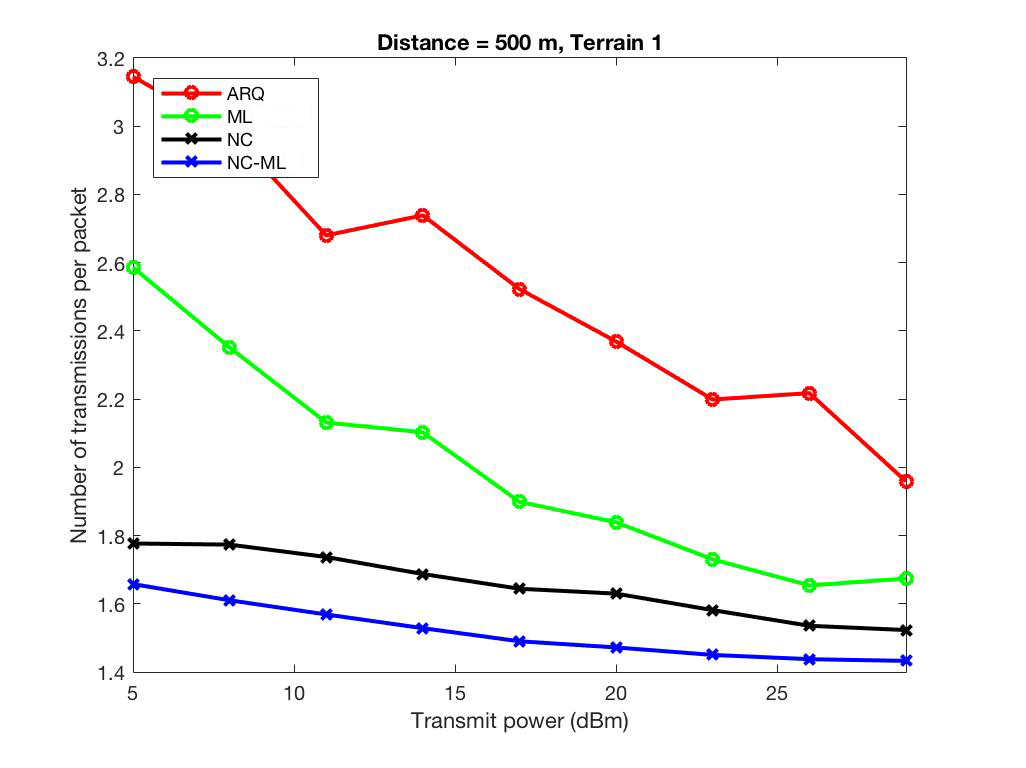}
\caption{Bandwidth efficiency vs. transmit power}
\label{efficiency_tran_power}
\end{figure}

Figure \ref{efficiency_tran_power} compares the bandwidth efficiency of different transmission schemes versus transmit power. In this simulation, we keep the distance between the transmitter and receivers at 500 meters and used the Terrain Type 1 as our transmission environment. As we can see, NC-ML achieves the best performance thanks to the machine learning algorithm that correctly predicts states of transmitted data packets. Consequently, it reduces the number of retransmissions compared with other schemes.

\begin{figure}[!t]
\centering
\includegraphics[width=2.7in]{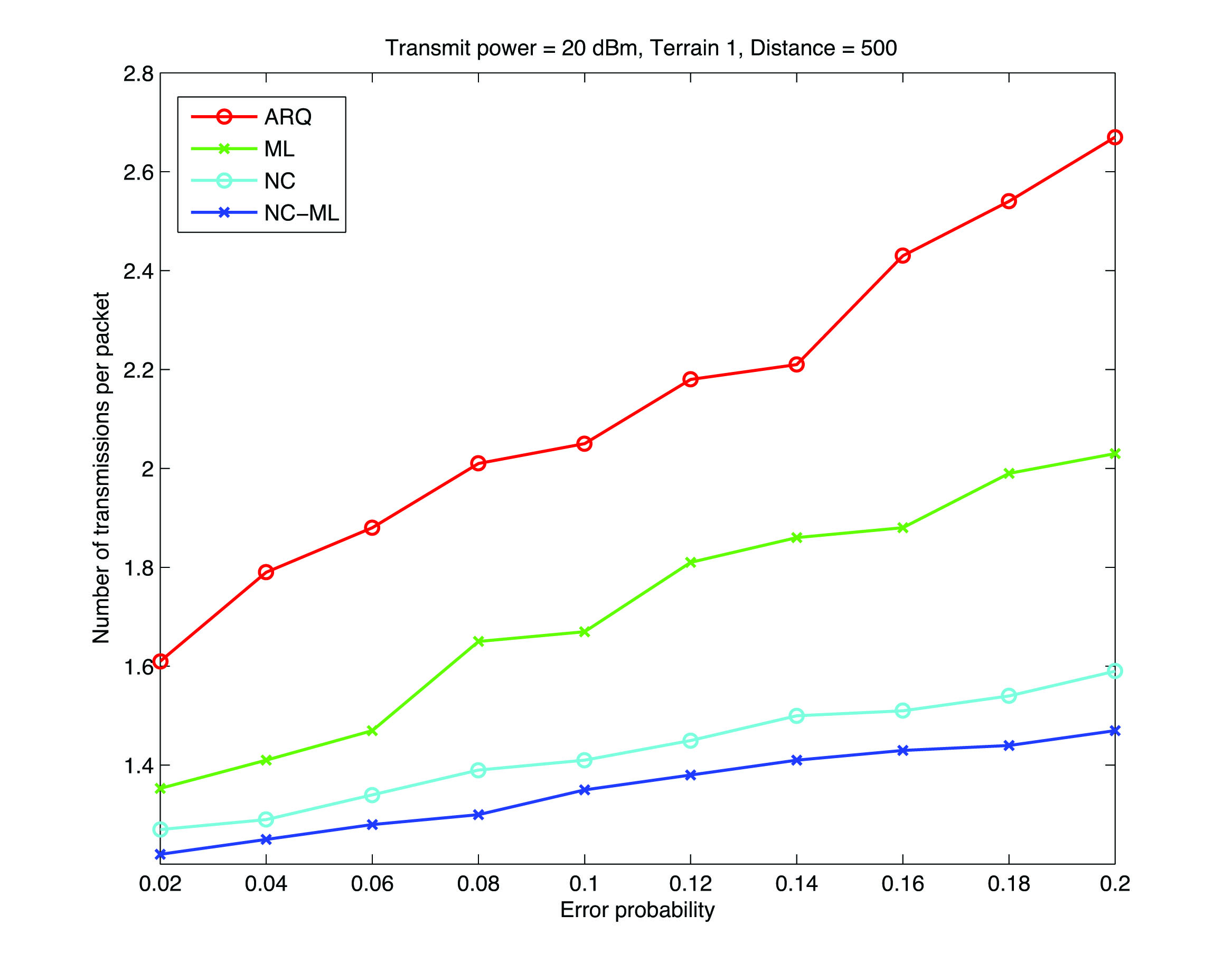}
\caption{Bandwidth efficiency vs. error probability of forward channels}
\label{efficiency_p_tran}
\vspace{-0.15in}
\end{figure}

\begin{figure}[!t]
\centering
\includegraphics[width=2.7in]{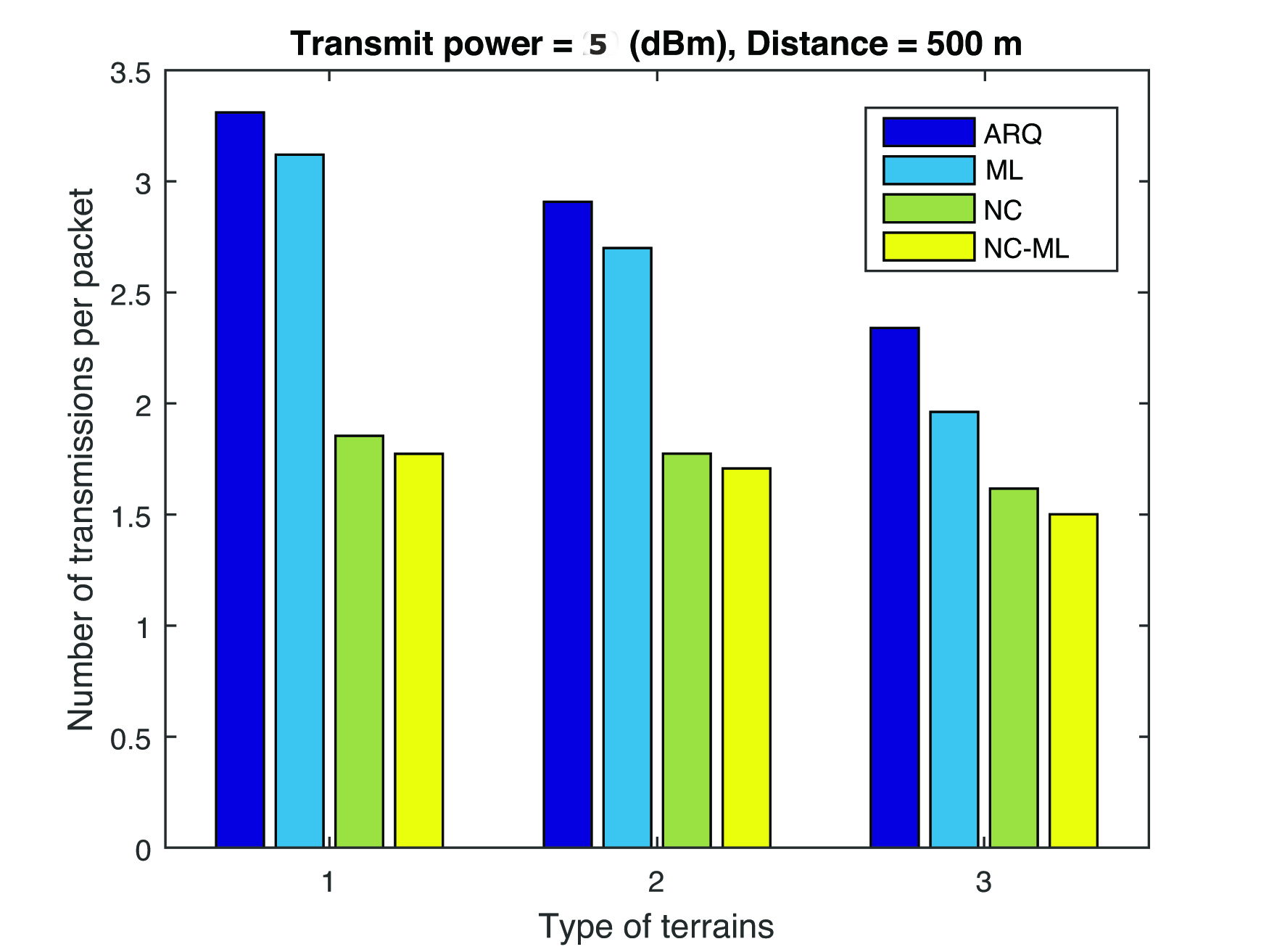}
\caption{Bandwidth efficiency vs. different types of terrains}
\label{efficency_terrain}
\vspace{-0.15in}
\end{figure}

Next, we evaluate the impact of forward channel conditions on the bandwidth efficiency of different schemes. In this simulation, we assumed the two receivers have the same packet error rate $p$ and varied it from 2$\%$ to 20$\%$. As $p$ increases, more packets will be corrupted at the receivers, as such, the number of transmission per packet increases. ARQ requires more number of retransmissions due to large number of lost packets. In addition, ARQ  suffers due to failing to estimate the actual states of the transmitted data packets. As a result, it consumes more bandwidth for retransmitting several unnecessary redundant packets. On the other hand, NC-ML exploits the advantages of both machine learning algorithm for predicting states of transmitted data packets and network coding for mixing data for retransmission, it significantly improves network bandwidth efficiency.

Finally, we compare network bandwidth efficiency of different schemes in different transmission environments in Figure \ref{efficency_terrain}. Particularly, we considered three types of terrains as specified in \cite{erceg1999} in our simulation. As expected the number of transmission per packet is largest in terrain type 1 while smallest in terrain type 3. This is because terrain type 1 models for hilly and high tree densities while terrain type 3 models for  flat terrain with light tree densities. In all terrain categories, NC-ML achieves the best performance thanks to the combination of machine learning algorithm and network coding for retransmission.
\vspace{-0.1in}
\section{Conclusion}
\label{sec:con}
In this paper, we have proposed a new approach by using machine learning algorithms in conjunction with network coding for efficient data transmission in lossy wireless networks. Our proposed machine learning algorithms are efficient and accurate by exploiting important features of network channels via historical feedback signals at the transmitter. Our simulation results show that the transmitter can accurately classify more than 90$\%$ states of transmitted data packets at the receivers. In combination with network coding for data retransmission, the proposed scheme NC-ML significantly increases the bandwidth efficiency compared with the existing approaches. Our future work will implement a testbed of the proposed approach.


\bibliographystyle{IEEEtran}
\bibliography{references}

\end{document}